# Enhanced Resolution of Poly-(Methyl Methacrylate) Electron Resist by Thermal Processing

Nima Arjmandi [a)], Liesbet Lagae, Gustaaf Borghs

IMEC, 75 Kapeldreef, 3001 Heverlee, Belgium

a)    Electronic Mail: Arjmandi@IMEC.be

Granular nanostructure of electron beam resist had limited the ultimate resolution of electron beam lithography. We report a thermal process to achieve a uniform and homogeneous amorphous thin film of poly methyl methacrylate electron resist. This thermal process consists of a short time-high temperature backing process in addition to precisely optimized development process conditions. Using this novel process, we patterned arrays of holes in a metal film with diameter smaller than 5nm. In addition, line edge roughness and surface roughness of the resist reduced to 1nm and 100pm respectively.

## I. Introduction

Electron beam lithography (EBL) is a high resolution, flexible, available and compatible lithography method. It has been used for high precision mask making, mold making for nanoimprint and many other applications. In particular, pattering very small holes is important in fabrication of magnetic storages, sub wavelength optical devices, supper lattices, photonic band gap waveguides, nanopores…

In the recent years, very small structures have been fabricated using EBL on different resists like hydrogen silsesquioxane (HSQ) [1] and ZEP (Nippon Zeon Co.) [2]. Nevertheless, the smallest structures are patterned by; polymethyl methacrylate (PMMA) electron beam resist and direct write on sodium chloride [3], silicon dioxide[4] and lithium fluoride [5] films. Direct write techniques suffer from problems in transferring the pattern to another thin film, and from the need of high energy, high dose and long exposure time. Therefore, PMMA seems to be the most applicable high-resolution resist for EBL.

By using the PMMA as the electron beam resist; lines as thin as 5nm [6, 7, 8] and 10nm gaps [9] have been reported. Although 5nm pillars have been achieved by etching thicker pillars [10, 11, 12] and holes as small as 10nm diameter on PMMA are reported [20, 21].

The most important problem that had limited the resolution of PMMA was its granular structure [17, 18, 19] (Fig.1.a). Given the fact that all the PMMA molecules in the resist solution are not exactly the same length; granules can form as fallows: after the spin coating and during the backing process, bigger PMMA molecules fold on themselves, due to the attraction between the monomers on the same chain. Hence, end-to-end distance of the polymer reduces to much shorter than its length. Moreover, solvent evaporation reduces the radius of gyration.

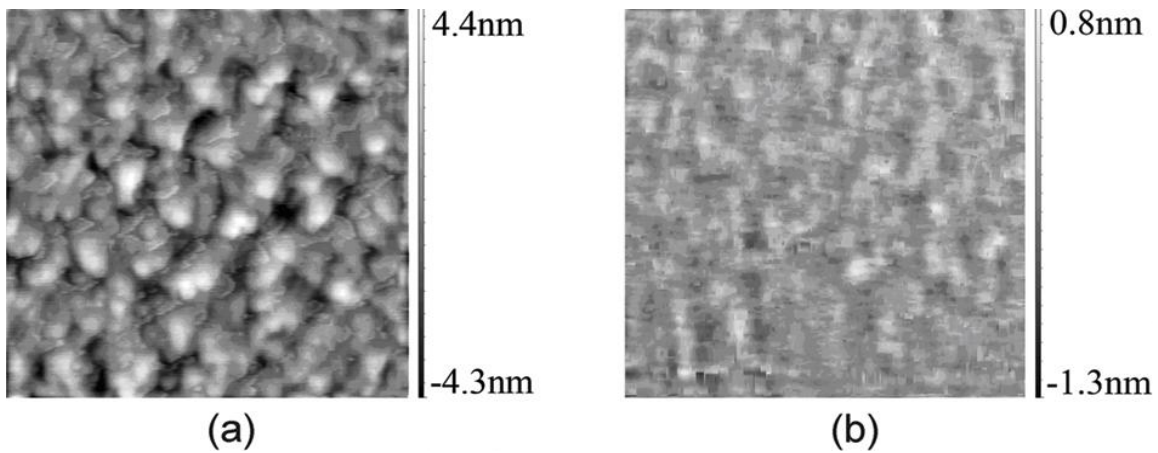

Fig. 1. 200x200nm AFM scans of PMMA after different baking processes. a) A typical AFM scan of sample that baked for a long time. In this case, it baked for 3 hours at 165°C. b) Baked at 250°C for 15sec while it was heated in 1sec and cooled in 1sec.

During the solvent evaporation, these big PMMA molecules immobilize sooner, while the smaller PMMA molecules are still highly mobile and because of the intermolecular attractions, these small molecules penetrate to the long molecules. This aggregation makes about 30nm big granules of PMMA. Because, the development is much faster at the granule boundaries, granules release one by one in to the developer instead of dissolving [19]. This effect increases the RMS of surface roughness to more than 250 pm [18] and the RMS of line edge roughness to more than 2nm and determines the smallest patternable structure on the PMMA. To solve this problem, we used a very short time thermal process, which does not give enough time to the short PMMA molecules to penetrate in to the long molecules, while using higher temperature to evaporate all the solvent in the limited time.

## II. Experiment

Low resistivity, p-type (100) silicon wafer were diced in 20X20mm pieces. Cleaning was performed by nitrogen spray fallowed by dipping in hot acetone, in hot isopropyl alcohol (IPA), in water and drying by nitrogen spray. Then, 10 nm ruthenium sputtered on the samples. To improve PMMA adhesion and uniformity; samples annealed at 400°C. Hereby, the contact angle of PMMA solution on the substrate's surface reduced to less than 5°. Samples were coated with 2% 950K PMMA in chlorobenzene spun at 6000rpm for 50sec with 1sec acceleration and 10sec deceleration time. Diverse thermal processes have been examined for baking the samples and the best results achieved by increasing the sample temperature from 21ºC to 250ºC in 1sec and keeping it in that temperature for about 15sec and cool it down to 21ºC in 1sec.

Unlike patterning lines, in patterning holes, there is a very confined space for developer to penetrate through the exposed area. Also, there is a very small space for dissolved PMMA to diffuse in to the bulk developer. To overcome the intermolecular force between unexposed walls and the exposed PMMA[8], also, to enhance the penetration of the developer in to the holes and also, to enhance the penetration of the dissolved PMMA in to the bulk developer, we used ultrasonic agitation. Other problems in patterning small holes were the gel thickness at the interface between the exposed PMMA and the developer (related to the radius of gyration) [13] and developer induced swelling [14]. To overcome these problems we used a weak developer, and cold development, which increases the contrast as well [15].

Electron beam exposure performed by Vistec VB_ 6HR with 50kev acceleration voltage and 150pA current. Development carried out in diverse conditions and the best condition has found to be 7:3 v:v IPA:water in ultrasonic bath at 12ºC. After the development, samples ion milled for about 65sec at 80mA beam current with 375v acceleration voltage and then, the resist residue striped by oxygen plasma. We have chosen ruthenium as the underlying layer; because of its high ion milling etch rate, high conductivity, small grain size, compatibility with different surfaces and the possibility to use it as a mask. In addition, ruthenium activates C-H and C-C bonds and benzene decompose on ruthenium at 87ºC, thus it will help the solvent to out gas during the baking, while there is no dehydrogenation bellow 277ºC. Thus, it will not damage the PMMA in the backing process, while PMMA has good adhesion on it.

SEM imaging performed by NOVA200 in vacuum and AFM imaging performed in atmospheric pressure in tapping mode with flexible tips to prevent PMMA deformation by the AFM tip.

## III. Results

Different thermal processes applied for baking and on each sample; line edge roughness (as defined in reference 19) and surface roughness measured by AFM (Fig. 1). An almost linear relation between the surface roughness and line edge roughness (Fig. 2.a) observed which substantiates a common source for both of them and the mentioned theory of PMMA granules[17, 18, 19]. In samples which have been baked at temperatures lower than 230°C, surface roughness and line edge roughness were improving by increasing the baking time (Fig.2. b). We believe it is due to reduction of solvent in the resist by evaporation. By baking in higher temperatures, an improvement in surface roughness and line edge roughness observed, when the baking time was very short, and by baking at 250°C for 15sec the lowest surface roughness and line edge roughness obtained which are about 100pm and 1nm respectively. We believe by reducing the baking time; PMMA molecules do not have enough time to diffuse and form the granules, while due to the high temperature all the solvent evaporates. By further increase of baking temperature, resist thinning and line edge rounding increases considerably, which is an indication of losing the contrast.

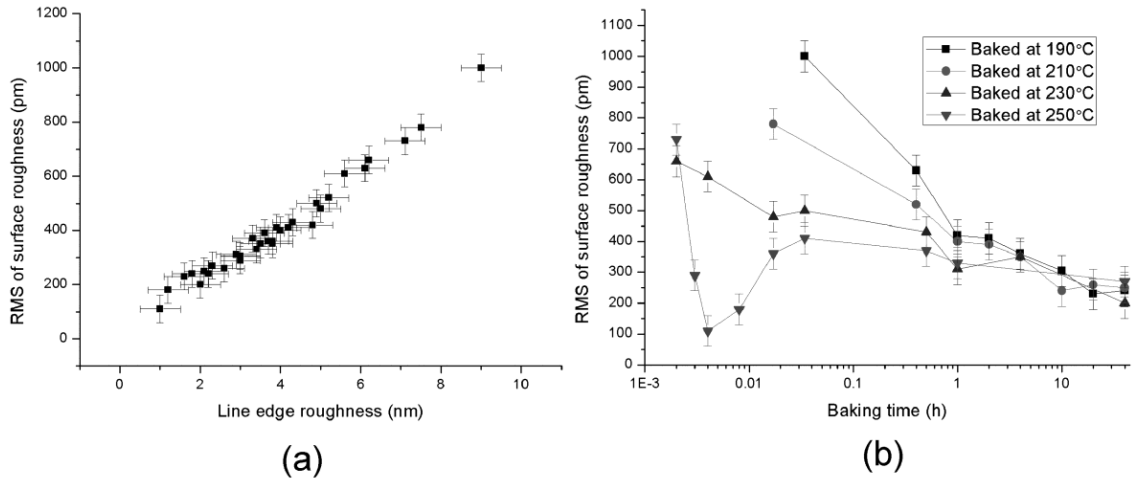

Fig. 2. a) RMS of surface roughness versus RMS of line edge roughness, obtained from different samples that prepared with different baking processes. b) RMS of surface roughness versus baking time.

In the samples, which were developed without ultrasonic agitation, small holes were not completely open, no matter for how long they were developed (an unopened hole could not result in a highly visible hole in SEM after ion milling). Developing in higher temperatures takes much shorter time but in higher or lower temperatures or stronger developers (like MIBK:IPA 3:1 v:v) contrast were much lower and pattern edge roughness was much higher that patterning small holes was not possible. But, by using the mentioned optimized development process; holes smaller than 5nm were obtained on the PMMA (Fig. 3). This higher resolution pattern transferred to the ruthenium film by ion milling and holes smaller than 5nm were obtained on ruthenium (Fig. 4).

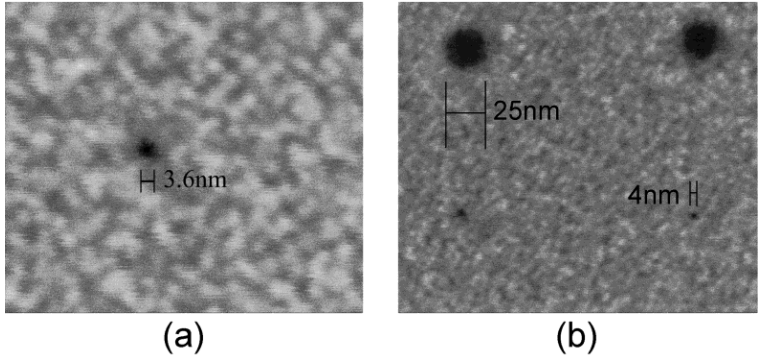

Fig. 3. SEM images of the holes patterned on PMMA (for SEM imaging; 1nm Pt has been sputtered on the samples). a) A high-resolution image of a smaller than 5nm hole. b) A part of hole arrays containing tow holes with 20nm diameter and tow holes with 5nm diameter.

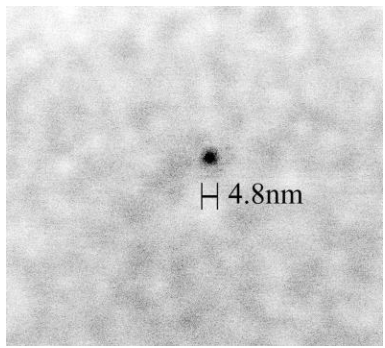

Fig. 4. A hole smaller than 5nm that has patterned in ruthenium.

## IV. Conclusion

Resolution of electron beam lithography is currently limited by the granular nanostructure of resists. To overcome this problem a very short time and high temperature backing process can be introduced, which results in a very smooth resist surface. By applying this baking process combined with the use of weak developer and ultrasonic agitation, holes smaller than 5nm were patterned in PMMA and the obtained pattern was successfully transferred to a metal thin film. As a result of this resist homogeneity; line edge roughness improved by a facture of two. Therefore, the resist nanostructure is currently not limiting the lithography resolution. By removing this limit, we are currently facing the intrinsic limits of the electron beam waist diameter as well as limitations in pattern transfer.

## V. References

[1] H. Namatsu et al., J. Vac. Sci. Technol. B 64, 69 (1998)

[2] K. Kurihara et al., Jpn. J. Appl. Phys., Part 1 34, 6940 (1995)

[3] A. N. Broers, IBM J. Res. Dev. 32, 502 (1998)

[4] A. N. Broers et al., Microelectron. Eng. 32, 131 (1996)

[5] W. Langheinrich et al., J. Vac. Sci. Technol. B 10, 2868 (1992)

[6] K. Yamazaki et al., Jpn. J. Appl. Phys. 43, 6B, 3767 (2004)

[7] S. Yasin et al., Appl. Phys. Lett. 78, 18, 2760 (2001)

[8] W. Chen et al., Appl. Phys. Lett. 62, 13, 1499 (1993)

[9] K. Liu et al., Appl. Phys. Lett. 80, 5, 865 (2001)

[10] W. Chen et al., J. Vac. Sci. Technol. B 11, 6, 2519 (1993)

[11] P. B. Fisher et al., J. Vac. Sci. Technol. B 11, 6, 2524 (1993)

[12] W. Chen et al., Appl. Phys. Lett. 63, 8, 1116 (1993)

[13] K. Ueberreiter, in diffusion in polymers, edited by J. Crank et al. (Academic, London, 1968), pp.219-257

[14] L. F. Thomson et al., J. Vac. Sci. Technol. 15, 938 (1978)

[15] W. Hu et al., J. Vac. Sci. Technol. B 22, 4, 1711 (2004)

[16] J. S. Greeneich, J. Appl. Phys. 45, 12, 5264 (1975)

[17] T. Ymaguchi et al., J. Photopolymer Technol. 10, 4, 635 (1997)

[18] E. A. Dobisz et al., J. Vac. Sci. Technol. B 15, 6, 2318 (1997)

[19] H. Namatsu et al., J. Vac. Sci. Technol. B 16, 6, 3315 (1998)

[20] O. Dial et al., J. Vac. Sci. Technol. B 16, 6, 3887 (1998)

[21] H. G. Craighead, J. Appl. Phys. 55, 12, 4430 (1984)